\documentclass{aa}
\usepackage{graphics}

\newcommand{\comment}[1]{}
        \def\smallskip{\vskip 2pt}

\begin{document}

\thesaurus{06(08.05.3; 08.06.3; 08.09.3; 08.12.1; 08.16.4; 08.14.2)}

\title{A unified formalism 
for the core mass-luminosity  relations of shell-burning stars}

\author{P. Marigo$^{1,2}$}
\institute{
$^1$ Dipartimento di Astronomia, Universit\`a di Padova,
        Vicolo dell'Osservatorio\ 5, I-35122 Padova, Italia \\
$^2$ Max-Planck-Institut f\"ur Astrophysik, Karl-Schwarzschild-Str.\
	1, D-87540 Garching bei M\"unchen, Deutschland
          }

\offprints{Paola Marigo \\ e-mail: marigo@pd.astro.it} 

\date{Received 11 April 2000 / Accepted 20 June 2000}

\maketitle
\markboth{Marigo P.}{The $M_{\rm c}-L$ relation}

\begin{abstract}

The luminosity evolution of
stars with highly condensed cores surrounded 
by nuclear-burning shell(s) is analytically investigated 
with the aid of homology relations.
With respect to earlier works using a similar approach (e.g.
Refsdal \& Weigert 1970; Kippenhahn 1981),  the major 
improvement is that 
we derive all the basic dependences
(i.e. on core mass, core radius, and chemical composition)  
in a completely generalised fashion, then
accounting for a large range of possible physical properties 
characterising the burning shell(s).
Parameterised formulas for the luminosity 
are given as a function of the
(i) relative contribution of the gas 
to the total pressure (gas plus radiation), (ii) 
opacity source, and (iii) dominant nuclear reaction rates.

In this way, the same formalism can be 
applied to shell-burning stars of various metallicities 
and in different evolutionary phases.
In particular, we present some applications concerning 
the luminosity evolution of  RGB and AGB stars with different 
chemical compositions, including the case of initial zero metallicity.
It turns out that homology predictions provide a good approximation 
to the results of stellar model calculations.

Therefore, the proposed formalism is useful to  
understand the possible differences in the luminosity evolution of 
shell-burning stars within a unified interpretative framework,
and can be as well adopted 
to improve the analytical description
of stellar properties in synthetic models. 
 

\end{abstract}


\section{Motivation of the work}
It is well known that 
the quiescent luminosity of a shell-burning star with degenerate core 
is essentially determined by its core mass, without
any dependence on the envelope mass, as extensively 
described in several works carried out in the past
(e.g. Eggleton 1967; Paczy\'nski 1970; Tuchman et al. 1983).
This property is usually referred to as the core mass-luminosity 
($M_{\rm c}-L$) relation, although other structural parameters
may affect the luminosity evolution, as indicated in the following.

Stellar evolutionary calculations 
actually confirm that
$M_{\rm c}-L$ relations are generally followed,   
for instance, by (i) low-mass stars during their
ascent on the Red Giant Branch (RGB) up
to the He-flash (e.g. Boothroyd \& Sackmann 1988); 
(ii) low- and intermediate-mass stars during the quiescent inter-pulse
periods of their Thermally Pulsing Asymptotic Giant Branch (TP-AGB)
evolution  
(e.g. Iben \&  Truran 1978, Wood \& Zarro 1981, Forestini \& Charbonnel 1997),
provided that hot-bottom burning is not operating
(see Bl\"ocker \& Sch\"onberner 1991); (iii) Planetary Nebula nuclei
as long as the H-burning shell is active (e.g. Vassiliadis \& Wood 1994);
and (iv) Nova systems during their stationary nuclear burning phases
(e.g. Tuchman \& Truran 1998).

These relations are known to be quite different for   
RGB and TP-AGB stars, and also vary among stars in the same
evolutionary stage but with different envelope chemical composition.
In view of interpreting these differences as the effect of
different physical conditions,
it would be advisable to derive quantitatively   
the dependence of the quiescent luminosity of a giant 
(RGB and TP-AGB) star on basic quantities, namely:
core mass $M_{\rm c}$, core radius $R_{\rm c}$, and 
chemical composition.

To this aim, we adopt the same formalism as fully described in Refsdal
\& Weigert (1970), which is based on the use of homology relations
applied to the case of stars with high-density cores surrounded by
nuclear burning shell(s).  
The authors demonstrated, for instance, that for low
values of the core mass (i.e. $M_{\rm c} \la 0.45 M_{\odot}$) and
negligible radiation pressure (i.e. $\beta \sim 1$) the luminosity is
expected to depend on the envelope mean molecular
weight as $L \propto \mu^{7 - 8}$.
Such theoretical prediction is found to describe extremely well the 
significant composition dependence of the luminosity of low-mass stars
evolving along the RGB, as shown by 
calculations of full stellar models (e.g. Boothroyd \& Sackmann 1988;  
see also Sect.~\ref{rgbzeta0}).
However, 
the results of Refsdal \& Weigert (1970) cannot be straightforwardly 
extended to the
case of stars evolving through the AGB phase, since the condition
$\beta \sim 1$ is generally not fulfilled, due to the increasing
importance of radiation pressure.

Therefore, our target is to generalize the formalism developed by
Refsdal \& Weigert (1970) for any value of $\beta$ in the range
$(0,1)$. To do this, we follow the indications suggested by Kippenhahn
(1981), who first pointed out that the quite different   
$M_{\rm c}-L$ relations for RGB and AGB stars are
indeed the expression of a unique relation modulated by the variation
of $\beta$. However, as Kippenhahn (1981) explicitly
derived the dependence of the luminosity 
only on $M_{\rm c}$ and $R_{\rm c}$, in this work we will extend the 
same kind of analysis to the composition dependence 
as a function of $\beta$.
In this way, we can predict to which extent the chemical properties
of AGB stars (with $\beta < 1$) 
may affect their $M_{\rm c}-L$ relation 
(see Sects.~\ref{agbchem}, and \ref{agbzeta0}). 

Moreover,
given the generality of the method,  
we can also investigate the sensitiveness of the $M_{\rm c}-L$ 
relation to the dominant nuclear energy source
operating the H-burning shell, i.e. the degree of temperature dependence
of the relevant reaction rates.
This will turn out to be significant in view of interpreting the  
peculiar luminosity evolution of RGB stars with initial zero-metallicity,
where the p-p chain (not the CNO cycle as is usually the case) 
 provides most of the 
stellar energy (see Sect.~\ref{rgbzeta0}). 
  
Finally, we invite the reader to
refer to the works by Refsdal \& Weigert (1970) and Kippenhahn (1981) 
for a full understanding of the analytical derivation, since many details will
be omitted here to avoid redundant repetitions and lengthy demonstrations.
However, the basic steps will be indicated in the next sections.

\section{The analytical method based on homology relations}
\label{teorcomp}
We will describe in the following 
the analytical procedure adopted to derive the dependence of the 
luminosity on $M_{\rm c}$, $R_{\rm c}$, and chemical composition.

Concerning this latter it can be seen,
from the basic stellar structure equations, that
the effect of the composition enters the $M_{\rm c}-L$ relation 
via three parameters (Refsdal \& Weigert 1970; Tuchman et al. 1983):
\begin{enumerate}
\item the mean molecular weight $\mu$;
\item the product of the abundances of the interacting nuclei involved
in the nuclear burning of hydrogen, $\epsilon_{0} = X^{2}$ (for the p-p
chain; $X$ is the hydrogen abundance in mass fraction) 
or $\epsilon_{0} = X Z_{\rm CNO}$ (for the CNO-cycle; $Z_{\rm CNO}$ is
the total abundance of CNO elements);
\item a factor $\kappa_{0}$ expressing the composition dependence of
the opacity $\kappa$.
\end{enumerate}  
The mean molecular weight 
 $\mu$ explicitly appears only in the equation of state:
\begin{equation}
\label{eqst}
P = P_{\rm G} + P_{\rm R} \propto \frac{\rho T}{\beta \mu}
\end{equation}
assuming that the total pressure, $P$,  is the 
sum of the contributions of the gas (supposed perfect), $P_{\rm G}$, and of
the radiation, $P_{\rm R}$. 
The quantity $\beta$ is defined as the ratio $P_{\rm G}/P$.
For a fully ionized gas $\mu = 4/ (5 X + 3 -Z)$.

The parameter $\epsilon_{0}$ is related to the rate of energy
generation by nuclear burning $\epsilon$, which can be 
conveniently approximated as:
\begin{equation}
\label{eps}
\epsilon = \epsilon_{0} \rho^{n-1} T^{\nu}
\end{equation}
with $n$ and $\nu$ being determined by the rates of the nuclear
reactions under consideration.
Typical values are: 
($n=2$, $\nu \sim 4$) for the p-p chain,
($n=2$, $\nu \sim 14 \div 20$) for the CNO-cycle, 
($n=3$, $\nu \sim 22$) for the triple $\alpha$-reaction.

The parameter $\kappa_{0}$ is related to the opacity, which can be 
expressed:
\begin{equation}
\label{kappa}
\kappa = \kappa_{0} P^{a}T^{b}
\end{equation}
with the exponents $a$ and $b$ depending on the dominating opacity source.
Note that in the case $\kappa$ is mostly due to the Thomson
electron scattering ($\kappa = 0.2(1+X)$ without any dependence on pressure 
and temperature), we get $\kappa_{0} = 1 + X$, and $a=b=0$.

Then,  
adopting the homology relations presented in
Refsdal \& Weigert (1970) we can express the luminosity
as a power-law relation:
\begin{equation}
\label{deptot}
L \propto M_{\rm c}^{\delta_1} R_{\rm c} ^{\delta_2} 
\mu^{\delta_{3}} \kappa_{0}^{\delta_{4}} \epsilon_{0}^{\delta_{5}} 
\end{equation}
which, under the assumptions of a fully ionized gas, 
with electron scattering opacity,  can be written:
\begin{equation}
\label{comptotcno}
L \propto M_{\rm c}^{\delta_1} R_{\rm c} ^{\delta_2} 
\left(\frac{4}{5X + 3 -Z}\right)^{\delta_{3}} 
(1+X)^{\delta_{4}} (X Z_{\rm cno})^{\delta_{5}} 
\end{equation}
for dominanting CNO-cycle, or 
\begin{equation}
\label{comptotpp}
L \propto M_{\rm c}^{\delta_1} R_{\rm c} ^{\delta_2} 
\left(\frac{4}{5X + 3 -Z}\right)^{\delta_{3}} 
(1+X)^{\delta_{4}} (X^2)^{\delta_{5}} 
\end{equation} 
for dominating p-p chain.

The exponents 
($\delta_{i}$, $i=1,5$) are the unknown quantities to be determined, 
as a function of $\beta$, of the opacity parameters ($a,\,b$), and of the 
energy parameters ($n,\, \nu$).

The composition dependence is the first being considered here,
for its particular relevance to the applications discussed in the
second part of this work (Sect.~\ref{appl}).  
For the sake of completeness, the
results for  
the dependences on  $M_{\rm c}$ and $R_{\rm c}$ are then briefly
presented (Sect.~\ref{teormcrc}). 

\subsection{The dependence on $\mu$}
\label{teormu}
The dependence on $\mu$ can be derived treating it as an independent
parameter to express the homology  relations:
\begin{eqnarray}
\rho(r/R_{\rm c}) & \propto & \mu^{\alpha_{3}} \nonumber \\
T(r/R_{\rm c})    & \propto & \mu^{\beta_{3}} \nonumber \\
\label{eq1}
P(r/R_{\rm c})    & \propto & \mu^{\gamma_{3}} \\
L(r/R_{\rm c})    & \propto & \mu^{\delta_{3}} \nonumber 
\end{eqnarray}
Such expressions imply the assumption that $\mu/\mu'=$
const. at each corresponding point (i.e. $r/R_{\rm c} = r'/R'_{\rm c}$) 
of models with different core radii (i.e. $R_{\rm c}$ and $R'_{\rm
c}$). Here $r$ denotes the radial coordinate of any point inside
the region extending from the bottom of the burning shell, $r=R_{\rm c}$,
up to a point, $r=r_{0}$, where the variables $\rho$, $P$, and $T$ 
have already significantly decreased, and where $L_{r} = L$.
Moreover, it is assumed that the other quantities (i.e.
$\rho$, $T$, $P$, $L$, $\epsilon_{0}$, and $\kappa_{0}$) do not vary
among models at corresponding points.

The exponents $\alpha_{3}$, $\beta_{3}$, $\gamma_{3}$, and $\delta_{3}$
are the unknown
parameters to be singled out.
To this aim, we need to integrate the equations of hydrostatic equilibrium,
radiative transport, and energy generation, basing on the
proportionality relations given in Eqs.~(\ref{eq1}), and using the 
expressions of Eqs.~(\ref{eps}) and (\ref{kappa}) for the rate of nuclear
energy generation and opacity, respectively. 
We then derive:
\begin{eqnarray}
P(r/R_{\rm c}) & \propto & \mu^{\alpha_{3}} \nonumber \\
\label{eq2}
T^{4-b}(r/R_{\rm c}) & \propto & \mu^{\alpha_{3} + a \gamma_{3} + \delta_{3}} \\
L(r/R_{\rm c}) & \propto & \mu^{n\alpha_{3} + \nu \beta_{3}} \nonumber 
\end{eqnarray}
Since there are four  unknown quantities, 
one more relation is needed in order to close the system of
Eqs.~(\ref{eq2}).  
This is given by the equation of state (Eq.~(\ref{eqst})),
which should be expressed in a more suitable form fully expliciting its
dependence on $\beta$.
For this purpose, we must consider that $\beta$
is a function of density, temperature, {\it and} molecular weight.
We can then derive the  three dependences as follows.     
As indicated by Kippenhahn (1981) the logarithmic derivatives of $\beta$
with respect to the density and temperature are:
\begin{equation}
\left(\frac{\partial \ln\beta  }{\partial \ln\rho}\right)_{T,\mu} = 1-\beta,
\,\,\,\,\,\,\,\left(\frac{\partial \ln\beta}{\partial \ln T}\right)_{\rho,\mu} = -3(1-\beta)
\end{equation}    
Similarly, we can make
a step ahead and derive also the logarithmic derivative of $\beta$ with
respect to the mean molecular weight:
\begin{equation}
\left(\frac{\partial \ln\beta  }{\partial \ln\mu}\right)_{T,\rho} =
\beta - 1
\end{equation}
Hence, the equation of state can be re-written:
\begin{equation}
P \propto \mu^{-\beta} \rho^{\beta} T^{4-3\beta}
\label{eqp}
\end{equation}
Note that in the limiting cases $\beta = 1$ and $\beta = 0$, we obtain
the right thermodynamical dependence of the total pressure when  
due to the sole contribution of gas and radiation, respectively.
It should also be remarked that Eq.~(\ref{eqp}) is expected to apply
for all $r/R_{\rm c}$. However,
since pressure changes by several order of magnitudes throughout the 
shell and radiative buffer above, it is far from obvious that 
Eq.~(\ref{eqp}) does apply throughout the region.
In fact it does, for two reasons: (1) in RGB stars because $\beta \sim 1$, 
and (2) in AGB stars because $\beta$ is constant throughout the shell
and sub-convective layers (``radiative zero'' condition).

At this point, all the necessary information is available.
Comparing the exponents of Eqs.~(\ref{eq1}), (\ref{eq2}), (\ref{eqp}),
we can solve the system of 4 algebraic equations
in the unknowns $\alpha_{3}$, $\beta_{3}$, $\gamma_{3}$, and
$\delta_{3}$, yielding:
\begin{equation}
\alpha_{3} = - \frac{\beta(\nu+b-4)}{(\nu+b-4)(1-\beta)+(1+a+n)(4-3\beta)}
\end{equation}

\begin{equation}
\beta_{3} = \frac{\beta(1+a+n)}{(\nu+b-4)(1-\beta)+(1+a+n)(4-3\beta)}
\end{equation}

\begin{equation}
\gamma_{3} = \alpha_{3}
\end{equation}

\begin{equation}
\label{del3}
\delta_{3} = 
\frac{\beta[\nu (1+a+n) -n(\nu+b-4)]}{(\nu+b-4)(1-\beta)+(1+a+n)(4-3\beta)}
\end{equation}
It is worth noticing that for $\beta = 1$ we get exactly the same
results as given by Refsdal \& Weigert (1970) (see their Eqs.~(27) and
Sect.~II d)), as expected.
 
\begin{figure*}
\begin{minipage}{0.80\textwidth}
\resizebox{\hsize}{!}{\includegraphics{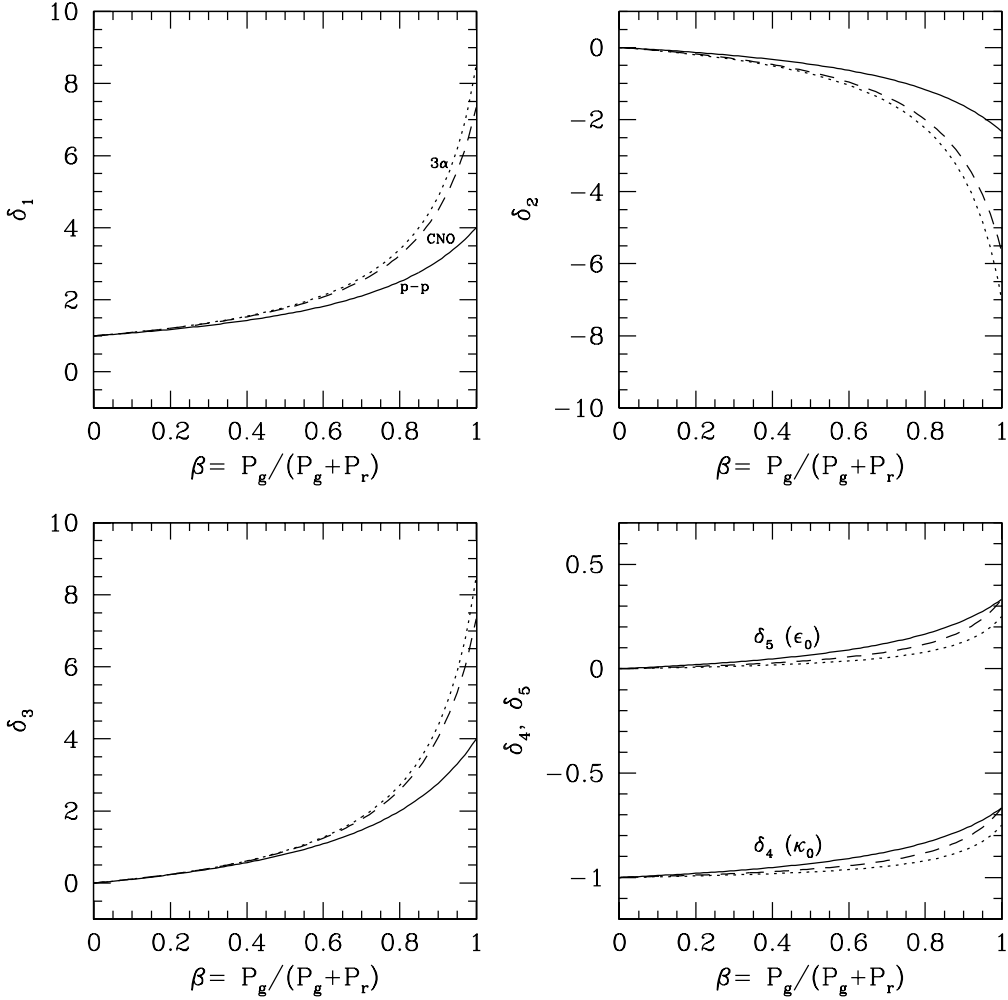}}
\end{minipage}
\hfill
\begin{minipage}{0.18\textwidth}
\caption{Expected behaviour, as a function of 
$\beta$, of the exponents in Eq.~(\protect\ref{deptot}): 
$\delta_{1}$ (related to the core mass),  
$\delta_{2}$ (related to the core radius),
$\delta_{3}$ (related to the mean molecular weight),
$\delta_{4}$ (related to the opacity),
and $\delta_{5}$ (related to the nuclear reaction rates).
We adopt $a=b=0$ for all cases, with 
($n=2$, $\nu = 4$) for the p-p chain (solid
line), ($n=2$, $\nu = 14$) for the CNO-cycle (dashed line), 
and ($n=3$, $\nu = 22$) for the triple $\alpha$-reaction (dotted line).}
\protect\label{d12345}
\end{minipage}
\end{figure*}

More generally,  
Eq.~(\ref{del3}) allows us to investigate the $\mu$-dependence of
the luminosity in the whole range $0 \le \beta \le 1$, so that it is
possible to set quantitative predictions both
for RGB stars (with $\beta \sim1 $) and AGB stars 
(with $\beta < 1$).
The results are displayed  in Fig.~\ref{d12345}, 
assuming $a=b=0$, and characteristic values of the  
the parameters ($n$, $\nu$) corresponding to 
relevant kinds of energy sources,
as already indicated in Sect.~\ref{teorcomp}.

From the inspection of 
the bottom-left panel of Fig.~\ref{d12345} 
the following features should be
noticed:
\begin{itemize}
\item In the case $\beta =1$ and dominant CNO cycle,
the exponent $\delta_{3} \sim 7$, which
well reproduces  the $\mu$-dependence of the luminosity for
RGB stars  as indicated by calculations of stellar models
(Boothroyd \& Sackmann 1988);
\item In the case $\beta =1$ and dominant p-p chain,
we get $\delta_{3} \sim 4$, i.e. a weaker $\mu$-dependence;  
\item In any case, for $\beta \sim 0$  the $\mu$-dependence
vanishes as expected when the gas pressure goes to zero;
\item In the case $\beta \sim 0.5$ -- $0.8$, which are typical
values found in low-mass AGB stars, and dominating
CNO-cycle, the $\mu$-dependence is not at all
negligible.
Specifically, it results  
$\delta_{3}\sim 1 \div 3$, a range 
 which  is consistent with the finding by Boothroyd \& Sackmann (1988)
(i.e. $L\propto \mu^{3}$)
in  their analysis of low-mass (hence with larger $\beta$) AGB stars.
\end{itemize}   

\subsection{The dependence on $\epsilon_{0}$ and $\kappa_{0}$}
\label{teorke}
In an analogous way as described in the previous section,
we aim now at deriving the dependence of the luminosity on both 
$\epsilon_{0}$ and  $\kappa_{0}$, already defined in
Sect.~\ref{teorcomp}.
We first write the homology relations:
\begin{eqnarray}
\rho(r/R_{\rm c}) & \propto & \kappa_{0}^{\alpha_{4}}\epsilon_{0}^{\alpha_{5}}  \nonumber \\
T(r/R_{\rm c})    & \propto & \kappa_{0}^{\beta_{4}} \epsilon_{0}^{\beta_{5}}  \nonumber \\
\label{eqek}
P(r/R_{\rm c})    & \propto & \kappa_{0}^{\gamma_{4}} \epsilon_{0}^{\gamma_{5}}  \\
L(r/R_{\rm c})    & \propto & \kappa_{0}^{\delta_{4}} \epsilon_{0}^{\delta_{5}} \nonumber 
\end{eqnarray}
assuming that the functions $\kappa_{0}(r/R_{\rm c})$ and
$\epsilon_{0}(r/R_{\rm c})$ scale up by a constant factor 
at corresponding points of models with
different core radii.

Then, integrating the basic stellar equations, expressing the total pressure
$P$ as in Eq.~(\ref{eqp}), comparing the exponents relative to the 
four variables $\rho$, $T$, $P$ and $L$, and solving the system of 8
algebraic equations, we finally  derive the 8 parameters:
\begin{eqnarray}
\alpha_{4} & = & 
\frac{(4-3\beta)}{(4-b-\nu)(1-\beta)-(1+a+n)(4-3\beta)} \\
\alpha_{5} & = & \gamma_{4} = \gamma_{5} = \alpha_{4}  \\
\beta_{4} & = & \beta_{5} =  
\frac{(1-\beta)}{(4-b-\nu)(1-\beta)-(1+a+n)(4-3\beta)} \\
\label{del4} 
\delta_{4} & = & \frac{n(4-3\beta)+\nu(1-\beta)}
                      {(4-b-\nu)(1-\beta)-(1+a+n)(4-3\beta)}\\
\label{del5}
\delta_{5} & = & 1 + \frac{n(4-3\beta)+\nu(1-\beta)}
                 {(4-b-\nu)(1-\beta)-(1+a+n)(4-3\beta)} 
\end{eqnarray}
Again, for $\beta=1$ the above equations yield the same results as in
Refsdal \& Weigert (1970), but now accounting for
any possible value of $\beta$.

Concerning the exponents related to the luminosity, $\delta_{4}$ and
$\delta_{5}$, their predicted trends are illustrated in 
the bottom-right panel of Fig.~\ref{d12345}.
We can notice that:
\begin{itemize}
\item their effect on the luminosity is usually  much weaker than that
produced by $\mu$, except for values of $\beta$ very close to zero; 
\item in the relevant range of $\beta$ for AGB stars, the predicted 
values for $\delta_{5}$ are in a rather 
good agreement with the results by 
Boothroyd \& Sackmann (1988), who quoted $L \propto Z_{\rm
cno}^{0.04}$ basing on their evolutionary calculations of AGB models;
\item in the case $\beta \sim 0$ and $a=b=0$ we get $\delta_{4} \sim -1$
and $\delta_{5} \sim 0$;
\item both $\delta_{4}$ and $\delta_{5}$ do not vary significantly 
adopting different values of the parameters $a$, $b$, $\nu$, and $n$,
within a reasonable range. 
\end{itemize}

\subsection{The dependences on $M_{\rm c}$ and $R_{\rm c}$}
\label{teormcrc}
In a similar fashion,
we can also derive the dependences of the luminosity on 
$M_{\rm c}$ and $R_{\rm c}$
\begin{equation}
L \propto M_{\rm c}^{\delta_{1}} R_{\rm c}^{\delta_2}.
\label{mcrc}
\end{equation}
For the sake of conciseness, we report here only the results for
$\delta_1$, and $\delta_2$ (see also Fig.~\ref{d12345}, top-left 
and top-right panels, respectively):
\begin{equation}
\delta_{1} = \frac{n a (4-3\beta) - n (4-b) - \nu (1+a\beta)}
{(4-b-\nu)(1-\beta) - (1+a+n)(4-3\beta)} {\rm ,} 
\label{del1}
\end{equation}
and
\begin{equation}
\delta_2 = \frac{\beta(3n+3b+\nu-3) - (3+n)[b+a(4-3\beta)]}
{(4-b-\nu)(1-\beta) - (1+a+n)(4-3\beta)}
\end{equation}
omitting the formulas for 
($\alpha_{i}$, $\beta_{i}$, $\gamma_{i}$; $i=1,2$).
The above equations for
$\delta_1$ and $\delta_2$
recover exactly those presented by Kippenhahn (1981) setting $a=b=0$, and 
those derived by Refsdal \& Weigert (1970) with $\beta=1$.

In brief, we can outline the following:
\begin{itemize}
\item $\delta_1$ and $\delta_2$ show opposite trends with $\beta$.
However, it should be recalled that, though $M_{\rm c}$ and $R_{\rm c}$ 
are formally treated as independent parameters, an $M_{\rm c}-R_{\rm c}$
relation is expected to exist for 
highly condensed cores which may be assimilated 
to white dwarf structures (Chandrasekhar 1939; Refsdal \& Weigert 1970;
Tuchman et al. 1983).
Considering the homology dependences of the luminosity only on
$M_{\rm c}$ and $R_{\rm c}$, we get the differential equation
(see Eq.~15 in Kippenhahn 1981):
\begin{equation}
\frac{d \ln L}{d \ln M_{\rm c}} = \delta_1 + \delta_2 \times 
\frac{d \ln R_{\rm c}}{d \ln M_{\rm c}}
\label{rcmc}
\end{equation}
where $d \ln R_{\rm c}/d \ln M_{\rm c}$ is the slope of the
$M_{\rm c}-R_{\rm c}$ relation.
In other words, the quantity $\delta_2 \times d \ln R_{\rm c}/d \ln M_{\rm c}$
represents the dependence on the core mass via $R_{\rm c}(M_{\rm c})$.
This can be estimated from Eq.~(\ref{rcmc}), e.g. taking the right-hand side 
from the results of stellar models, 
and evaluating $\delta_1$ and $\delta_2$  from the homology relations.
\item As for $\mu$, the dependence on $M_{\rm c}$ grows  at increasing
$\beta$. For $\beta \sim 1$, we get typical relations 
$L \propto M_{\rm c}^7$ for dominant CNO-cycle, and  $L \propto M_{\rm c}^4$
for dominant p-p chain.
Again, the former prediction very well agrees with the results
of evolutionary stellar calculations of RGB stars (Boothroyd \& Sackmann 1988).
The effect of a weaker $M_{\rm c}$-dependence predicted in the latter case
will be investigated in Sect.~\ref{rgbzeta0}.
In both cases ($\delta_1 \sim 7$ and
$\delta_2 \sim 4$), the luminosity should be just moderately affected
by changes in the core radius, i.e. the product 
$|\delta_2 \times d \ln R_{\rm c} / d \ln M_{\rm c}|$ (calculated with
Eq.~(\ref{rcmc}))
is typically less than unity for $\beta \sim 1$.
\item At decreasing $\beta$ the dependence on $M_{\rm c}$
quickly approaches a linear one ($\delta_1=1$ for $\beta=0$),
whereas the dependence on $R_{\rm c}$ tends to vanish 
($\delta_2=0$ for $\beta=0$). 
This result is consistent with the typical flatter slopes ($\sim 1-2$) 
of the fitting $M_{\rm c}-L$ relations derived from evolutionary
calculations of AGB stars. 
\end{itemize}

\section{Some applications}
\label{appl}
\subsection{The  $M_{\rm c}$ -- $L$ relation on the RGB: 
the zero-metallicity case}
\label{rgbzeta0}
\begin{figure*}
\resizebox{\hsize}{!}{\includegraphics{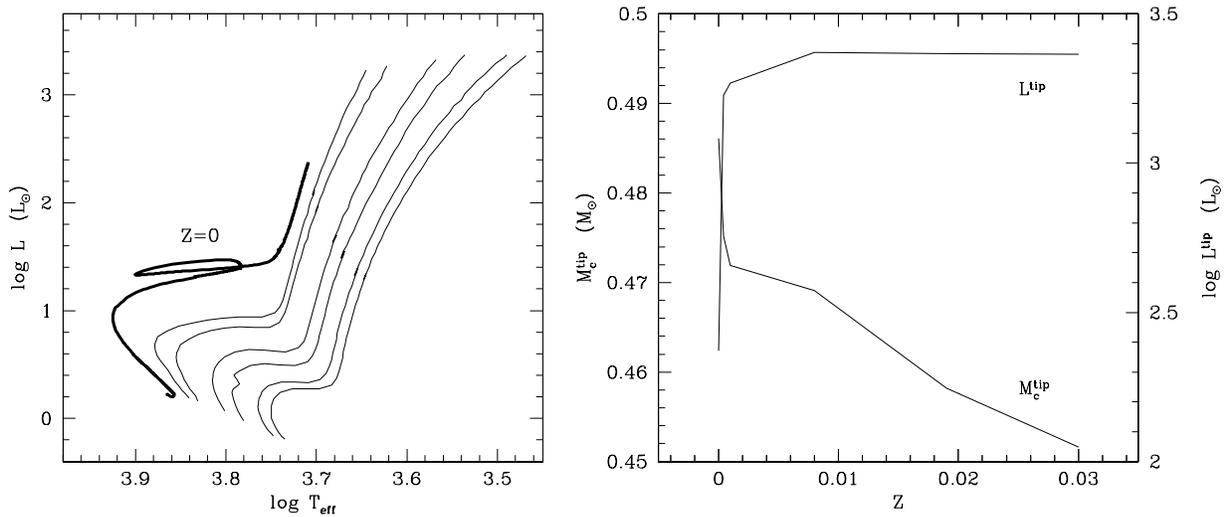}}
\hfill
\caption{Left-hand side panel: Evolutionary tracks  up to the He-flash
of 1 $M_{\odot}$ models
with different initial metallicities (i.e. $Z=$0.030, 0.019,0.008, 
0.004, 0.001, 0.0004 from right to left; taken from Girardi et al. 2000)
(thin lines), and zero metallicity (thick line; 
Marigo et al. 2000, in preparation).
Right-hand side panel: luminosity and core mass on the RGB tip of
1 $M_{\odot}$ models as a function of metallicity.}
\label{fig_hr_1}
\end{figure*}

\begin{figure*}
\begin{minipage}{0.67\textwidth}
\resizebox{\hsize}{!}{\includegraphics{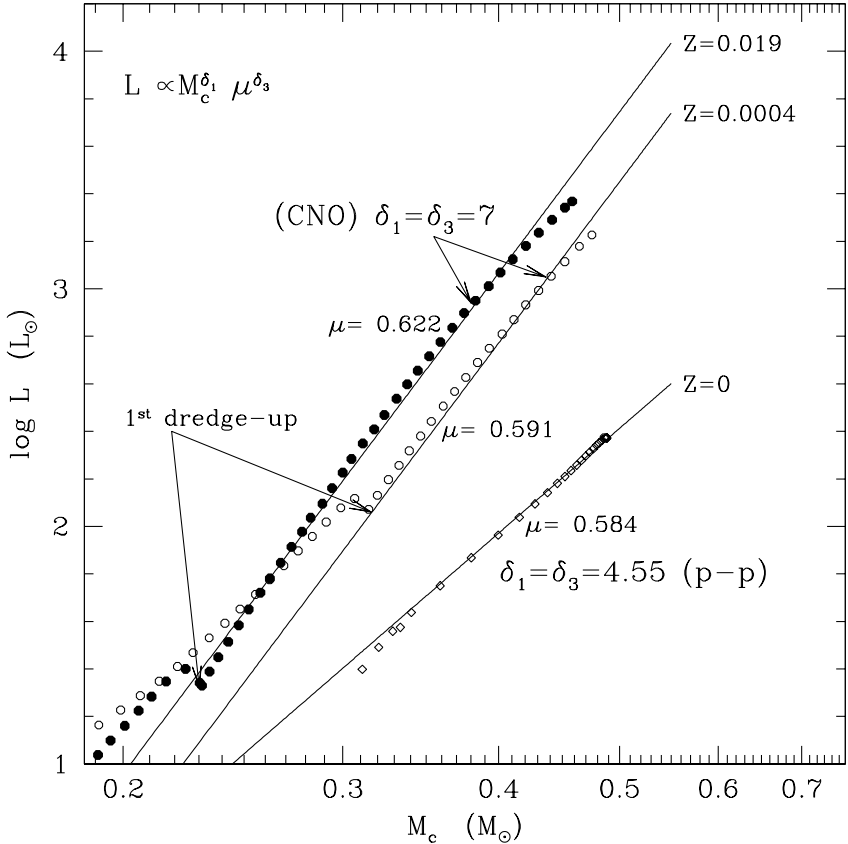}}
\end{minipage}
\hfill
\begin{minipage}{0.32\textwidth}
\caption{$M_{\rm c}-L$ relation on the RGB. 
Symbols correspond to selected
models for 1 $M_{\odot}$  stars with different metallicities, 
and mean molecular weights after the first dredge-up as indicated. 
Predictions from 
homology relations are shown as solid lines. See text for further
explanation.}
\protect\label{fig_rgblmc}
\end{minipage}
\end{figure*}

Evolutionary calculations of low-mass models 
(see Figs.~\ref{fig_hr_1} and \ref{fig_rgblmc}) 
indicate that 
\begin{enumerate}
\item towards lower metallicities and for given stellar mass 
the luminosity at the tip of the RGB  
decreases, whereas the core mass increases; 
\item at decreasing metallicity 
the luminosity on the RGB is lower for given core mass; and 
\item the slope of the $M_{\rm c}-L$ relation
on the RGB is flatter for $Z=0$ models than for $Z \neq 0$ models.
\end{enumerate}
The first point clearly confirms that the luminosity is not solely 
a function of the core mass.
To better interpret the above trends we can make use of the analytical
relations obtained in the previous sections.

Let us consider the $M_{\rm c}-L$ relation presented by Boothroyd
\& Sackmann (1988)
\begin{equation}
L = (11.6 M_{\rm c} \mu)^7 Z_{\rm cno}^{1/12}
\label{bs88} 
\end{equation}
which is a fitting formula of evolutionary calculations for RGB
stars with different metallicities (masses and luminosities are in solar
units).
 
The check the reliability of homology predictions, we first 
compute the exponents
$\delta_1$ and $\delta_3$ with the aid 
of Eqs.~(\ref{del1}) and (\ref{del3}), 
and compare them with those given in Eq.~(\ref{bs88}).
To do this, $\beta$, the opacity parameters, and the
energy parameters should be specified.
We set $\beta = 1$  -- which is proved to be a good approximation 
for all RGB models here considered --,  and $a=b=0$ 
-- for the sake of simplicity.
The adoption of more proper values for $a$ and $b$ would
result in very small corrections, as pointed out by Refsdal \& Weigert (1970).
%
%
%

The choice of the energy parameters requires some comments.
It turns out that the bulk of energy produced in the H-burning shell 
is provided by the CNO-cycle for both  $1 M_{\odot}$ RGB models with
$Z=0.019$ and $Z=0.0004$, whereas energy production 
is dominated by the p-p chain in the $1 M_{\odot}$ RGB 
model with $Z=0$.
It follows that typical values ($n=2$, $\nu = 14-16$) and ($n=2$, $\nu = 4-6$)
should be adopted in the two cases, yielding 
$\delta_{1} \sim \delta_{3} \sim 7-8$ and 
$\delta_1 \sim \delta_{3} \sim 4.0-4.7$ for RGB models
in which the CNO cycle and p-p chain dominates, respectively.

The predictions for the former case 
(CNO dominated) are in excellent agreement with 
Eq.~(\ref{bs88}), thus reproducing the results
of complete stellar calculations of RGB stars with dominating CNO cycle.
Moreover, it is worth remarking that
points (1) and (2) -- mentioned at the beginning of this section -- 
are explained as the effect of differences   
in the mean molecular weight, i.e. for given $M_{\rm c}$ and lower $\mu$, 
$L$ is lower.

The predictions for the latter case (p-p dominated) fully explain point (3).
In fact, the weaker temperature dependence 
of p-p reactions results in a flatter slope of the 
$\log M_{\rm c} -\log L$ relation followed by the 1 $M_{\odot}$ model
with initial zero metallicity.
A good fit is obtained adopting 
\begin{equation}
L = (11.6 M_{\rm c} \mu)^{4.55}
\end{equation}   
with $L$ and $M_{\rm c}$ in solar units.
This power-law relation 
has the same base as in Eq.~(\ref{bs88}), 
but a different exponent, derived under the
assumption of dominating p-p reactions with energy parameters
($n=2$, $\nu\sim 5.6$).
Taking for $\mu$ the value after the first dredge-up as indicated by 
the calculations by Girardi et al. (2000), and letting $M_{\rm c}$ vary
over the relevant range, we finally obtain the relation shown
in Fig.~\ref{fig_rgblmc},  
which remarkably well matches the evolutionary results at 
initial zero metallicity.

To summarise the conclusions, we can notice that, for $\beta \sim 1$,
 the {\sl mean molecular weight}  determines the luminosity level  
of the  $M_{\rm c}-L$ relation 
(i.e. the {\sl intercept} in logarithmic plot)
for RGB models 
with similar energy properties,
whereas the  kind of {\sl nuclear energy source} affects 
the rate of the geometric increment of the luminosity  
with the core mass (i.e. the {\sl slope}
 of the $\log M_{\rm c} - \log L$ relation).
The latter point explains the fact that the 
$\log M_{\rm c} - \log L$ relations of the $1 M_{\odot}$ models with
$Z=0.019$ and $Z=0.0004$ are almost parallel, whereas the relation
for $Z=0$ runs flatter.
  
%
\begin{figure*}
\begin{minipage}{0.80\textwidth}
\resizebox{\hsize}{!}{\includegraphics{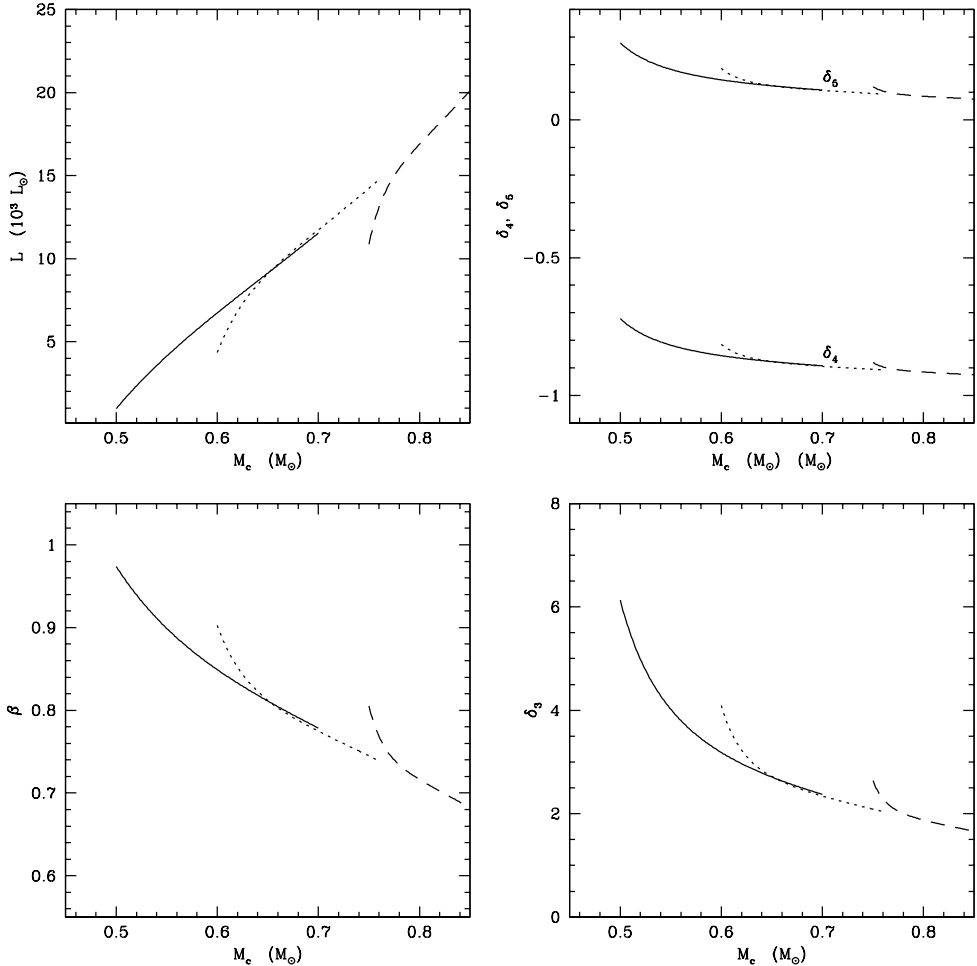}}
\end{minipage}
\hfill
\begin{minipage}{0.18\textwidth}
\caption{Predicted evolution as a function of the core mass 
of the pre-flash maximum luminosity $L$, ratio of the gas pressure
over total pressure $\beta$ estimated with Eq.~(\ref{beta}),  
and exponents $\delta_{3}$, $\delta_{4}$, and $\delta_{5}$, 
from the first thermal pulse of three TP-AGB models.}
\protect\label{beta_wag}
\end{minipage}
\end{figure*} 
%
%

%
%
\begin{figure*}
\begin{minipage}{0.67\textwidth}
\resizebox{\hsize}{!}{\includegraphics{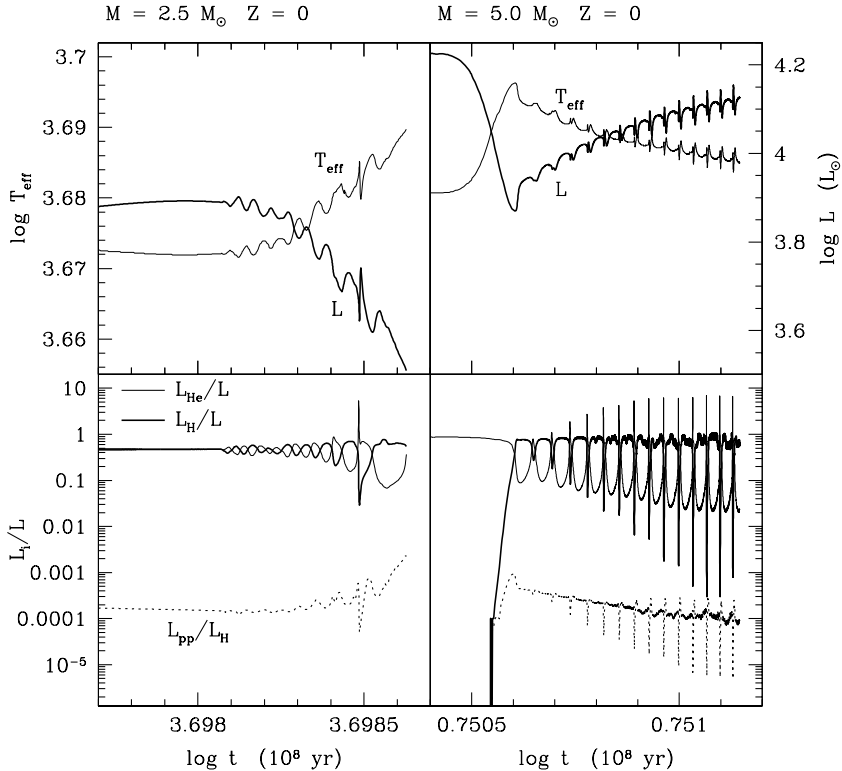}}
\hfill
\end{minipage}
\begin{minipage}{0.32\textwidth}
\caption{Surface and energy properties of AGB models with
initial zero metallicity and masses of 2.5 $M_{\odot}$ 
(left-hand side  panels), and 5.0 $M_{\odot}$ (right-hand side panels),
since the first appearance of the He-shell thermal instabilities.
The fractional contributions of shell nuclear burnings to the surface
luminosity ($L_{\rm H}/L$ and $L_{\rm He}/L$) are displayed, together
with that provided by p-p chain relative to the integrated energy generation
from shell H-burning ($L_{\rm pp}/L_{\rm H}$)}.
\protect\label{fig_agbzeta0}
\end{minipage}
\end{figure*} 
%

%
\begin{figure*}
\begin{minipage}{0.67\textwidth}
\resizebox{\hsize}{!}{\includegraphics{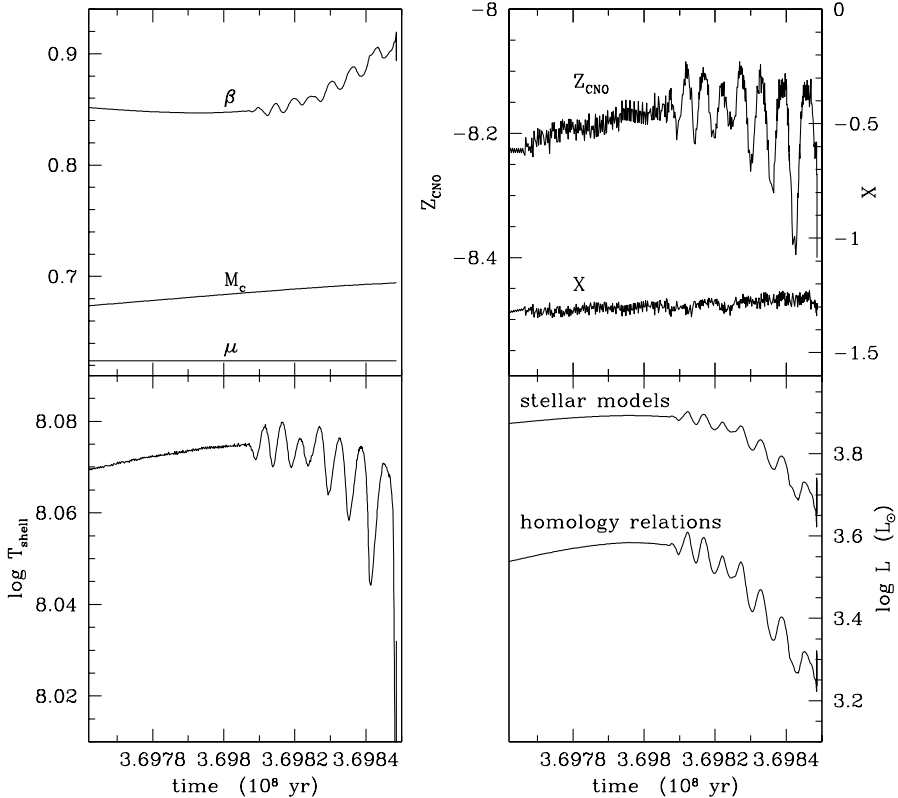}}
\hfill
\end{minipage}
\begin{minipage}{0.32\textwidth}
\caption{Time behaviour of relevant quantities characterising the 
evolution of 2.5 $M_{\odot}$ AGB model, namely:
the gas pressure over total pressure ratio ($\beta$), and
the mass coordinate ($M_{\rm c}$) at the top of the H-burning shell;
the mean molecular weight in the envelope ($\mu$); the abundances (by number) 
of CNO elements ($Z_{\rm CNO}$) and hydrogen 
($X$), and the temperature ($T_{\rm shell}$)
at the point of maximum energy production
in the H-burning shell. The ''true'' stellar luminosity predicted
by full stellar models is compared to that obtained from homology
relations (arbitrarily shifted by a constant on the logarithmic axis).}
\protect\label{fig_agb2p5}
\end{minipage}
\end{figure*} 
%

\begin{figure*}
\begin{minipage}{0.67\textwidth}
\resizebox{\hsize}{!}{\includegraphics{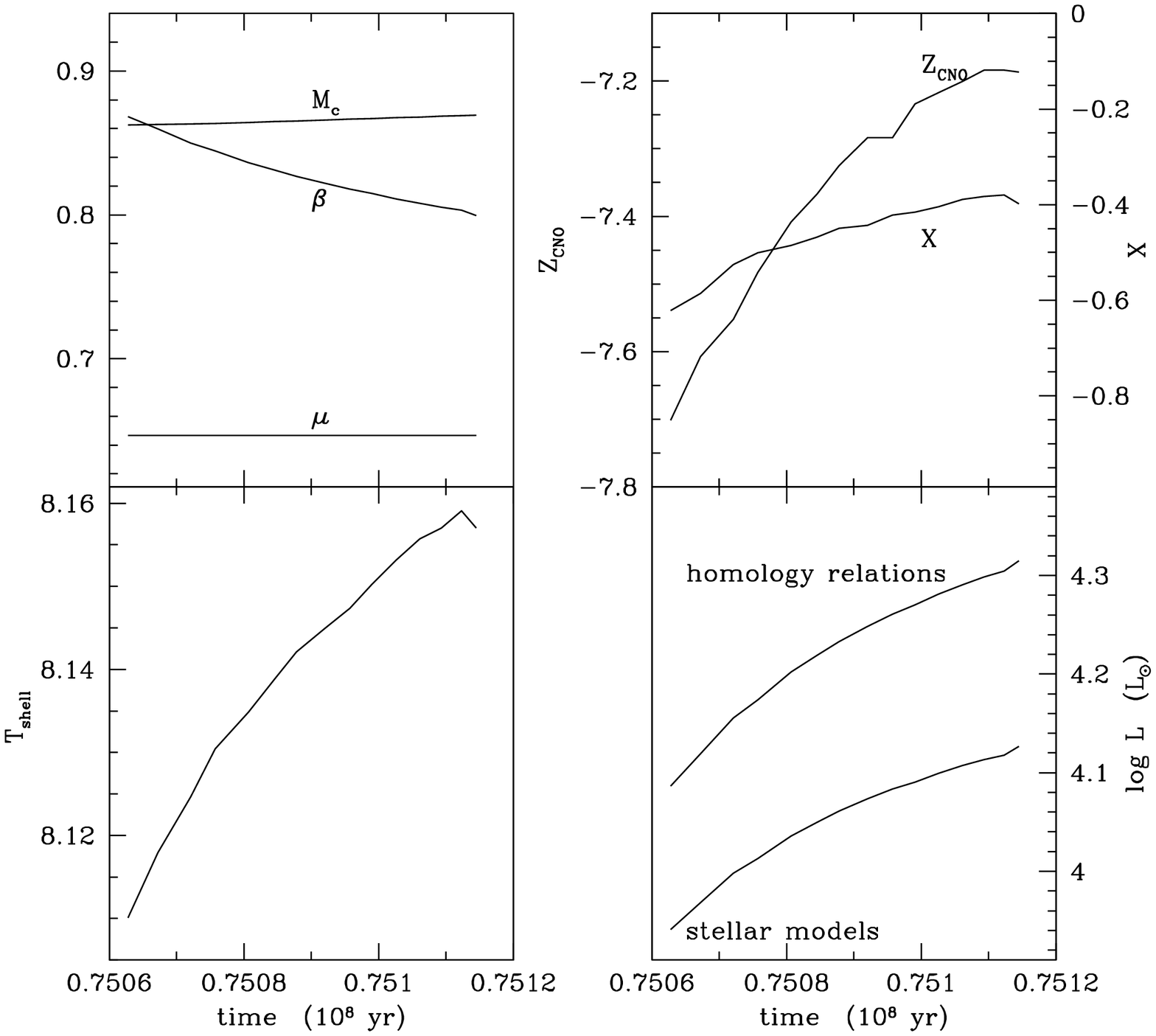}}
\hfill
\end{minipage}
\begin{minipage}{0.32\textwidth}
\caption{The same as in Fig.~\protect\ref{fig_agb2p5}, but for
the (5 $M_{\odot}$, $Z=0$) model. The quantities are shown at the
stage of the maximum quiescent luminosity immediately preceding a thermal
pulse.}
\protect\label{fig_agb5}
\end{minipage}
\end{figure*} 

\subsection{The  composition dependence of $M_{\rm c}-L$ relation 
on the TP-AGB phase}
\label{agbchem}
Let us now apply the relations derived in Sects. \ref{teormu} and 
\ref{teorke}  to TP-AGB stars, 
in order to estimate the degree of dependence  
of the quiescent luminosity on envelope composition. 

First of all, we need to know how $\beta$ 
varies during the TP-AGB evolution of a stellar model, so that
we can directly obtain $\delta_{3}$ from Eq.~(\ref{del3}).
To this aim, we make use of the definition of radiative gradient:
\begin{equation}
\nabla_{\rm r} = \frac{3}{16 \pi a c G} \kappa \frac{L_{r}}{M_{r}} 
\frac{P_{r}}{T_{r}^{4}}
\end{equation}
with usual meaning of the quantities.
In the context of the present study, the radial coordinate $r$ 
refers to the top of the H-burning shell or, equivalently, to some
point near 
the bottom of the
radiative inert region (extending up to the base of the convective envelope).
Here $L_{r}/M_{r}$ is nearly constant 
and equal to $L/M_{\rm c}$, the opacity is dominated by Thomson
electron scattering, i.e. $\kappa = 0.2 (X+1)$, and the radiative
gradient, $\nabla_{\rm r}$, 
approaches the ``radiative zero'' value of 0.25 
(see, for instance, Scalo et al. 1975).
Under these conditions, and reminding that 
$P_{r}/T_{r}^{4} \propto (1 - \beta)^{-1}$, we can express $\beta$
as a function of the core mass $M_{\rm c}$, surface luminosity $L$ 
and hydrogen abundance $X$ in the envelope:
\begin{equation}
\label{beta}
\beta =  1 - 7.956 \times 10^{-6} \,\, (1+X) 
\frac{L/L_{\odot}}{M_{\rm c}/M_{\odot}}.
\end{equation}
Figure \ref{beta_wag} illustrates the results of synthetic 
calculations for three TP-AGB models with original solar composition.
The evolution of the quiescent luminosity (top-left panel)
is computed according to  the prescription
presented by  Wagenhuber \& Groenewegen (1998).
This gives an accurate fit to the results of extensive 
full evolutionary calculations of the TP-AGB phase (Wagenhuber 1996), 
expressing  the quiescent luminosity as a function
of the core mass, from the first thermal pulse.

Then, once $\beta$ is evaluated with the aid of Eq.~(\ref{beta})
for current values of $L$ and $M_{\rm c}$ (top-left panel),
the composition parameters $\delta_{3}$ (Eq.~\ref{del3}), 
$\delta_{4}$ (Eq.~\ref{del4}), and $\delta_{5}$ (Eq.~\ref{del5})
can be calculated (right panels).
In other words, we are able to predict the current composition dependence  
of the quiescent luminosity as the star evolves on the TP-AGB.

These simple synthetic calculations 
are meant to be indicative examples, showing
the expected sensitiveness of the luminosity to possible changes 
in the surface chemical composition during the evolution caused, 
for instance, by convective dredge-up episodes.
For the sake of simplicity, these events are assumed not to occur 
in the cases under consideration, to avoid
the complication due to feed-back effects. 
We will discuss this point below in
this section.

An interesting point to be noticed in Fig.~\ref{beta_wag} is that
the composition dependence is quite strong during the initial 
(and fainter) part of the TP-AGB phase 
 (e.g. $\delta_{3} \sim 5-6$ in the first
stages of the model with the lowest core mass), then 
becoming weaker as the full-amplitude regime is attained.
This trend  simply reflects the rate of increase of the 
radiation pressure (i.e. $\beta$ decreases) during the evolution
(bottom-left panel of Fig.~\ref{beta_wag}).   
Therefore,  we would expect that,  
if an efficient dredge-up occurs quite early during the  
TP-AGB evolution, the increase of the mean molecular weight  could 
alter the asymptotic approach towards the full amplitude 
regime, as  otherwise expected for unchanged chemical composition.
As already suggested by Marigo  et al. (1999), 
this prediction,
in combination with additional effects, 
may concur to explain the recent results by Herwig et
al. (1998), who find 
a steeper increase of the luminosity 
in stellar models with extremely efficient dredge-up (with $\lambda \sim
1$ and larger) already from the first thermal pulses.

Another point to be remarked is that any  
change in  the envelope  composition is expected to produce
a certain feed-back on the luminosity, in addition to the
direct effect already discussed.  
In fact, a variation $\Delta \mu > 0$ produces $\Delta L > 0$,
since $\delta_{3}$ is always positive (we do not consider here 
the extreme case  $\beta = 0$ for which $\delta_{3}=0$).  
At the same time, we get $\delta T > 0$ and $\delta P < 0$
(as $\beta_{3}> 0$ and $\gamma_{3} < 0$), both causing 
$\delta \beta < 0$ and $\Delta L < 0$.
In other words, the effective  increase of $L$ due to an increment of
$\mu$ is somewhat reduced with respect to that directly predicted by
Eq.~(\ref{del3}).
Similar effects are produced by variations of the other two composition
parameters, $\epsilon_{0}$, and $\kappa_{0}$ (see also Sect. II g) 
in Refsdal \& Weigert 1970).

In conclusion, as already suggested by 
Kippenhahn (1981) and confirmed by stellar 
evolutionary calculations (see, for instance Boothroyd \& Sackmann 1988), 
we quantitatively demonstrate that 
the composition dependence of the 
luminosity of TP-AGB stars is weaker than for RGB stars (due
to the increasing importance of the radiation pressure).
However, non-negligible effects on the luminosity can be driven by significant
variations of the mean molecular weight, either related to originally different
chemical compositions, or caused by the third dredge-up.
The occurrence of the latter process already since  the first thermal
pulses  could significantly affect the luminosity evolution, in particular,
of low-mass TP-AGB stars (with large $\beta$). 
Finally, we remark that the relations 
presented in Sects.~\ref{teormu} and \ref{teorke} can be usefully employed
to improve the analytical description of the 
luminosity evolution  in synthetic AGB models.
\subsection{The case of zero-metallicity AGB stars}
\label{agbzeta0}
The last application refers to the TP-AGB phase 
of stars with initial zero metallicity.
In Fig.~\ref{fig_agbzeta0} we show the time evolution of the surface properties
($L$ and $T_{\rm eff}$) and of the contributions 
of the nuclear energy sources (He- and H-burning
shells) for two AGB models of different
masses (2.5 $M_{\odot}$, and 5.0 $M_{\odot}$), taken 
from Marigo et al. (2000, in preparation). 

As we can see from Fig.~\ref{fig_agbzeta0}, 
the  2.5 $M_{\odot}$ experiences  weak luminosity fluctuations instead
of ``normal'' thermal pulses, 
whereas the 5 $M_{\odot}$ model shows the occurrence of rather 
strong He-shell flashes.
Actually, it has 
already been pointed out by Sujimoto et al. (1984; see also 
Chieffi \& Tornamb\`e 1984,  and Dom\`\i nguez et al. 1999) that 
the occurrence of thermal pulses
in zero-metallicity AGB stars  is critically  dependent on the
core mass and abundance of CNO elements in the envelope.
To this respect, a detailed discussion
of our zero-metallicity models is given in Marigo et al. (2000, in preparation) and will not be repeated here.
  
What we simply aim to do in this work is  
to test whether the homology predictions presented 
in Sect.~\ref{teorcomp} are able to account for 
the quite different trends in the surface properties
of the two AGB models here considered. In fact, as
we can notice from the top panels of  Fig.~\ref{fig_agbzeta0}, 
the 5 $M_{\odot}$ model
is climbing the AGB 
at increasing luminosities (and decreasing effective temperatures), 
whereas the 2.5 $M_{\odot}$ model is clearly
evolving downward on its Hayashi track.
Moreover, in both cases the p-p chain is negligibly contributing
to the nuclear energy generated within the H-shell 
($L_{\rm pp}/L_{\rm H} \la 10^{-4}$), i.e. the CNO-cycle is the dominant
energy source. 
We also report that the third dredge-up is never found to occur 
in these models
up to the moment at which the calculations were stopped. 

Once $\beta$ is estimated with the aid of Eq.~(\ref{beta}),
the exponents $\delta_{3}$, $\delta_{4}$, and $\delta_{5}$ 
can be calculated assuming $a=b=0$, and 
($n=2$, $\nu=14$), the latter being suitable for dominant CNO cycle.
The luminosity is then  derived from Eq.~(\ref{comptotcno}), taking the 
values of the core mass and  composition factors 
($\mu$, $X$, and $Z_{\rm CNO}$) from selected models of the 2.5 $M_{\odot}$
and 5 $M_{\odot}$ stars (see Figs.~\ref{fig_agb2p5} and \ref{fig_agb5}).
It turns out that in both cases the trend in
the luminosity evolution predicted by stellar evolutionary
calculations is satisfactorily reproduced by homology relations.

In particular, the oscillating behaviour in  the surface luminosity 
of the 2.5 $M_{\odot}$ model shows up in correspondence with
that of $\beta$, which presents a mirror-like trend.
The average increase of $\beta$ 
actually determines the long-term decrease of the luminosity, despite
of the progressive increment of the core mass.
Moreover, it is interesting  to notice that, 
in the regime of oscillating luminosity, the total abundance of CNO elements
-- at the point of maximum nuclear energy efficiency
in the H-burning shell -- also presents clear fluctuations, 
reflecting a similar
trend in the temperature at the same mesh-point.
To this regard, it should be remarked that 
although for this model the CNO abundance
in the envelope is zero (as it is not changed by any dredge-up episode),
the CNO catalysts in the H-burning shell are self-produced,  
starting from the synthesis of primary $^{12}$C 
via the triple $\alpha$-reaction,
operating at typical shell temperatures.

The increase with time of the pre-flash maximum luminosity
in the 5 $M_{\odot}$ model is also well reproduced by  
homology predictions, being essentially determined by the rate 
of increase of the core mass. Finally, we can notice that, contrary to 
the 2.5 $M_{\odot}$ model, in this case $\beta$ is decreasing. 
 
\section{Concluding remarks}
A {\it general formalism} based on homology relations is presented 
to derive the structural dependences of the quiescent 
luminosity of shell-burning stars 
in different evolutionary phases (RGB and AGB), with 
different chemical compositions, and different nuclear energy sources.

The reliability of this formalism is tested through several consistency
checks.
First, we are able to get exactly the 
same formulas as in earlier similar works for specific choices of the 
parameters, e.g. Refsdal \& Weigert (1970) for $\beta=1$.
Second, our predictions are found to be in good agreement with some 
basic results of complete stellar calculations.
In particular, 
as far as the composition
dependence of the $M_{\rm c}-L$ relation is concerned, 
we recover the finding of evolutionary 
calculations that $L \propto \mu^{7}$
for RGB stars with $Z_{\rm cno} > 0$, 
and $L \propto \mu^{3}$ for AGB stars with 
$0.5 M_{\odot} \la M_{\rm c} \la 0.7 M_{\odot}$ (e.g. Boothroyd 
\& Sackmann 1988).
Moreover, 
according to our results, the effect on the luminosity of TP-AGB stars 
produced by significant composition changes  
should be larger in the case of low-mass stars that experience the 
third dredge-up  after the first thermal pulses. 

The case of zero-metallicity giant stars 
is also investi\-gated.
We show that 
the particular luminosity evolution of RGB stars with 
$Z=0$ can be very well 
explained  by  considering 
not only the dependence on the 
mean molecular weight, but also the kind of dominant
nuclear energy source (i.e. the p-p reactions).
Moreover, a good reproduction of the luminosity trend of AGB models with 
initial zero metallicity is obtained.

Finally, it is worth remarking that the analytical prescriptions 
presented in this work could be usefully employed in synthetic 
evolution models to improve them in accuracy, and to test the 
effects of different physical conditions, i.e. 
dominant nuclear reaction rates, opacity source,  
relative contributions of gas/radiation to the total pressure,  
and chemical composition. 
%

\begin{acknowledgements}
I like to thank L. Girardi for his careful  
reading of the manuscript and useful comments, and the
anonymous referee for important remarks on this work.
\end{acknowledgements}

\end{document}